\documentclass[onecolumn]{aastex631}
\usepackage{graphicx}
\usepackage{epstopdf}
\usepackage{amssymb}
\usepackage{mathrsfs}
\usepackage{natbib}
\usepackage{longtable}
\usepackage{array}
\usepackage{rotating}

\begin{document}

\title{The physical mechanism for two rapid changing-look AGNs: SDSS J0225+0030 and SDSS J1723+5504}

\author[0000-0002-7299-4513]{Shuang-Liang Li}
\affiliation{Shanghai Astronomical Observatory, Chinese Academy of Sciences, 80 Nandan Road, Shanghai 200030, People's Republic of China}

\author[0000-0002-2355-3498]{Xinwu Cao}
\affiliation{Institute for Astronomy, School of Physics, Zhejiang University, 866 Yuhangtang Road, Hangzhou 310058, People's Republic of China}

\author[0000-0003-3137-1851]{Wei-Min Gu}
\affiliation{Department of Astronomy, Xiamen University, Xiamen, Fujian 361005, People’s Republic of China}

\correspondingauthor{Shuang-Liang Li, Xinwu Cao, Wei-Min Gu}
\email{lisl@shao.ac.cn, xwcao@zju.edu.cn, guwm@xmu.edu.cn}

\begin{abstract}
SDSS J0225+0030 and SDSS J1723+5504 are two turn-on changing-look active galactic nuclei (CL AGNs) with transition timescales shorter than one year. Such short timescales pose a challenge for the current physical models of CL AGNs. We investigate this issue by exploring two possible mechanisms in this work. First, we consider the effect of a large-scale magnetic field on the viscous timescale, which can increase the radial velocity of the accretion disk. However, it is found that the timescale given by this model remains significantly longer than one year. Second, we improve the model of \citet{2025ApJ...988..207L}, which proposed that the inner thin disk in the bright state may form through the collapse of an advection-dominated accretion flow (ADAF) in the dim state, rather than being replaced by the advection of the outer thin disk. We re-estimate the transition radius $R_{\rm tr}$ between the inner ADAF and the outer thin disk through the observed variation of optical flux between the bright state and dim state. It is found that $R_{\rm tr}$ can be significantly reduced in these two objects owing to the lower gas temperature in the inner disk region (of the order of $10^4$ K), resulting from their large black hole masses ($\sim 10^9 M_{\odot}$) and small Eddington-scaled mass accretion rates ($\sim 0.01$). The cooling timescales given by the revised model in these two objects are found to be comparable to the observed transition timescales.     
\end{abstract}


\section{Introduction}

Changing-look active galactic nuclei (CL AGNs) are a fascinating class of AGNs that exhibit quick state transitions between type 1 and type 2, characterized by the appearance or disappearance of broad emission lines on timescales ranging from several months to tens of years (e.g., \citealt{parker_detection_2016,gezari_iptf_2017,yang_discovery_2018,macleod_changing-look_2019,wang_x-ray_2020,2021A&A...650A..33P,2022ApJ...933..180G,2023NatAs...7.1282R,2024ApJS..272...13P,2025ApJ...986..160D,2025ApJS..278...28G,2025ApJ...980...91Y}). This short timescale challenges the AGN unified model \citep{antonucci_unified_1993,1995PASP..107..803U}, providing an excellent laboratory to explore both the physics of accretion processes and the dynamics of the AGN broad emission line region.

The physical mechanisms of CL AGNs remain under debate. The previously favored dust obscuration model has been found inadequate to explain the multi-wavelength properties of CL AGNs, as the observed strong variability in infrared emission and the low polarization in optical-UV band are inconsistent with its predictions \citep{2017ApJ...846L...7Sheng,ross_new_2018,hutsemekers_polarization_2019,2021A&A...650A..33P,2023ApJ...953...61Y}. Furthermore, the typical timescales of dust obscuration are too long compared with the observed timescales of CL AGNs (e.g., \citealt{2021A&A...650A..33P}). Tidal disruption events (TDE) have also been proposed as a possible mechanism for a small fraction of CL AGNs \citep{ricci_destruction_2020,2023MNRAS.526.2331C}, but can't account for other objects.
Currently, it is generally suggested that the variation of global mass accretion rate in the accretion disk of AGNs can trigger the changing-look of AGNs (e.g., \citealt{2023NatAs...7.1282R,2023ApJ...953...61Y}). 
However, one key issue remains unresolved in this model: the observational timescales in CL AGNs are far shorter than the viscous timescales associated with the variation of accretion rate in a thin disk \citep{shakura_black_1973}. Although the large-scale poloidal magnetic field threading on the accretion disk can help to solve this issue by transferring the disk angular momentum with outflow and enhancing the radial velocity of gas, leading to the decrease of viscous timescale \citep{2013ApJ...765..149Cao,2014ApJ...788...71L,2019ApJ...872..149L,2021ApJ...916...61F,2023ApJ...958..146W}. However, most CL AGNs are radio-quiet, suggesting that the combination of large-scale magnetic fields and black hole spin may not be sufficient to launch powerful jets in these sources.


In previous work, we investigated the physical mechanism of radio-quiet turn-on CL AGNs (\citealt{2025ApJ...988..207L}, hereinafter LC2025). The accretion flow of CL AGNs in the dim state (type 2) is suggested to be composed of an inner advection-dominated accretion flow (ADAF) and an outer thin disk \citep{2019ApJ...883...76R,2022ApJ...927..227L}. Since the viscous timescale of a thin disk is too long, we first proposed that the physical mechanism for radio-quiet turn-on CL AGNs may be related to the cooling timescale of an ADAF, i.e., the inner thin accretion disk is collapsed from an ADAF instead of being formed from the advection of an outer thin disk (LC2025).
By comparing the observed transition timescale of turn-on CL AGNs ($t_{\rm tran}$) with the cooling timescale of an ADAF ($t_{\rm cool}$) with the critical mass accretion rate (see LC2025 for details), it is found that $t_{\rm cool}$ is shorter than $t_{\rm tran}$ in most objects, validating the feasibility of our model. However, there are still two 'outlier' objects, i.e., SDSS J0225+0030 and SDSS J1723+5504, showing a much shorter $t_{\rm tran}$ than $t_{\rm cool}$. SDSS J0225+0030 is a CL quasar exhibiting both turn-off and turn-on behaviors at redshift 0.504, which is discovered by \citet{2016MNRAS.457..389M} based on the repeat photometry from Sloan Digital Sky Survey (SDSS) and the repeat spectra from SDSS-III Baryon Oscillation Spectroscopic Survey (BOSS). Its broad emission line (BEL, $\rm H_{\rm \beta}$) completely disappeared from MJD=52944 to 55445 and reappeared after 254 days in the rest frame (from MJD=55445 to 55827). 
\citet{2021A&A...650A..33P} reported another CL AGN (SDSS J1723+5504) with short transition timescale based on the SDSS DR7, which was found to transition from type 1.8 to type 1 in 142 days in the rest frame. In addition, SDSS J1723+5504 has only two observations, so the derived observed transition timescale is an upper limit. Although SDSS J0225+0030 has another observation between the bright and dim states (with a slight increase in the continuum flux at 5100 Angstrom, \citealt{2024ApJS..272...13P}), the resulting timescale is also an upper limit. If we could conduct higher-cadence spectroscopic observations on CL AGNs (e.g., \citealt{2025arXiv251018445D}), the inferred transition timescales may be further shortened.

Such short timescales (one order of magnitude smaller than the typical timescales of CL AGNs) pose a challenge to our current physical models. As mentioned above, the prevailing physical mechanisms for CL AGNs include mass accretion rate variation, dust obscuration and TDE. These scenarios can be further discriminated through their multi-band physical properties, such as continuum evolution, emission-line response, infrared echo, X-ray behavior, and radio properties. Though radio and X-ray data are unavailable in these two objects, we can still distinguish their physical mechanisms using the available optical spectroscopic and photometric data \citep{2016MNRAS.457..389M,2021A&A...650A..33P}. Firstly, the TDE scenario is disfavored for SDSS J0225+0030, as its light curve shows a clear "dimming and re-brightening" cycle rather than the characteristic $t^{-5/3}$ decay of a TDE. For SDSS J1723+5504, the narrow-line region diagnostics confirm that the gas has been ionized by an AGN-like continuum over the past $10^3-10^4$ years, which is inconsistent with a TDE scenario. Secondly, the difference spectra of both objects can be well reproduced by a standard thin disk model ($f_\nu \propto \nu^{1/3}$). This spectral shape is inconsistent with the simple dust extinction scenario, which would redden the continuum. Besides, the Mg II emission line in SDSS J0225+0030 responds linearly to the continuum change, which is consistent with reprocessing or intrinsic accretion changes, but not with simple dust obscuration. Furthermore, the timescales predicted by the dust obscuration model with reasonable parameters are significantly longer than those observed. These results indicate that the state transitions in these two objects are triggered by the variation of mass accretion rate in the accretion disk, especially by the transition in the accretion mode. 

In this work, we will further investigate the possible physical mechanism for these two objects. As a preliminary check, we first calculate the viscous timescales of a thin accretion disk with a large-scale magnetic field. While the magnetic field can indeed decrease the viscous timescales by transferring the angular momentum, we find that the timescales given by this model are still too long compared with the observed transition timescales. This suggests that employing a large-scale magnetic field in a thin disk is insufficient to account for the observed timescales (see the appendix for details). We therefore turn to another scenario: the revised LC2025 model in the following section.

\section{Model} \label{model}

The cooling timescale $t_{\rm cool}$ provided by the model of LC2025 is proportional to $\Omega_{\rm K}^{-1}$ ($\sim R^{3/2}$). Therefore, the accuracy of $t_{\rm cool}$ strongly depends on the precise estimation of the transition radius $R_{\rm tr}$ between the inner ADAF and outer thin disk. In LC2025, this value is roughly estimated by requiring that the temperature at the transition radius $R_{\rm tr}$ should be greater than the characteristic temperature of the disk region dominating the 5100 \AA{} emission ($\sim$ 5700 K), since the bolometric luminosity is calculated with the flux at 5100 \AA{} (see LC2025 for details).
Obviously, $R_{\rm tr}$ given in this method should be the maximum, indicating that the cooling timescale $t_{\rm cool}$ is also a maximum.

In this work, we improve the LC2025 model by enhancing the precision of $R_{\rm tr}$. The accretion disk in the dim state of CL AGNs is suggested to be composed of an inner ADAF and an outer thin disk, where the inner ADAF will gradually transition to a thin disk when the mass accretion rate slightly increases \citep{2019ApJ...883...76R,2022ApJ...927..227L}, leading to the enhancement of optical flux. Therefore, we can estimate the transition radius $R_{\rm tr}$ through the variation of optical flux between the bright state and dim state. This method is more physical and is promising for greatly decreasing the transition radius $R_{\rm tr}$. The optical flux at 5100 \AA{} in the rest frame in the bright state ($f_{\rm {\nu, 5100}}$) and that in the dim state ($f_{\rm {\nu, 5100}}^{'}$) can be given as
\begin{equation}
f_{\rm {\nu, 5100}}=\frac{4\pi h {\rm cos}i \nu^3}{c^2D^2} \int^{R\rm out}_{R_{\rm in}} \frac{R {d}R}{e^{h\nu/kT_{\rm eff}}-1},
\end{equation}
and
\begin{equation}
f_{\rm {\nu, 5100}}^{'}=\frac{4\pi h {\rm cos}i \nu^3}{c^2D^2}  \int^{R\rm out}_{R_{\rm tr}} \frac{R {d}R}{e^{h\nu/kT_{\rm eff}}-1}, 
\end{equation}
respectively, where $D$ and $i$ are the luminosity distance and the angle of inclination, respectively. 
The frequency $\nu$ is set to $\nu=5.88\times 10^{14}$ Hz (5100 \AA{}). For simplicity, we assume a Schwarzschild black hole with the inner radius $R_{\rm in}=3R_{\rm s}$ ($R_{\rm s}=2GM/c^2$) and the outer radius $R_{\rm out}=1000 R_{\rm s}$, respectively. $R_{\rm in}=3R_{\rm s}$ represents the inner radius of a thin disk in the bright state. The effective temperature of an accretion disk $T_{\rm eff}$ is given by:
\begin{equation}
    T_{\rm eff}=\left[\frac{3GM_{\rm bh}\dot{M}}{8\pi \sigma R^3}\left(1-\sqrt{\frac{3R_{\rm s}}{R}}\right)\right]^{1/4},
\end{equation}
where $\sigma$ and $G$ are the Stefan–Boltzmann
constant and the gravitational constant, respectively. $T_{\rm eff}$ can be rewritten as: 
\begin{equation}
T_{\rm eff} \sim 6.5*10^7 m_{\rm bh}^{-1/4} \dot{m}^{1/4} (R/R_{\rm s})^{-3/4} {\rm K}, 
\end{equation}
where $m_{\rm bh}=M_{\rm BH}/M_{\odot}$ ($M_{\rm BH}$ and $M_{\odot}$ are the black hole mass and solar mass, respectively). The Eddington-scaled mass accretion rate $\dot{m}$ is given by $\dot{m}=\dot{M}/\dot{M}_{\rm Edd}$, where $\dot{M}$ and $\dot{M}_{\rm Edd} = L_{\rm Edd}/0.1c^2$ are the mass accretion rate and the Eddington mass accretion rate, respectively. For these two objects, the temperature at $R_{\rm out} = 1000 R_{\rm s}$ is only $\sim 650$ K ($m_{\rm bh}\sim 10^9$, $\dot{m}\sim 10^{-2}$), contributing little to the optical flux. Therefore, it is safe to adopt $R_{\rm out}=1000 R_{\rm s}$ as the outer radius.

\section{Results} \label{results}

\begin{figure}
\centering
\includegraphics[width=12cm]{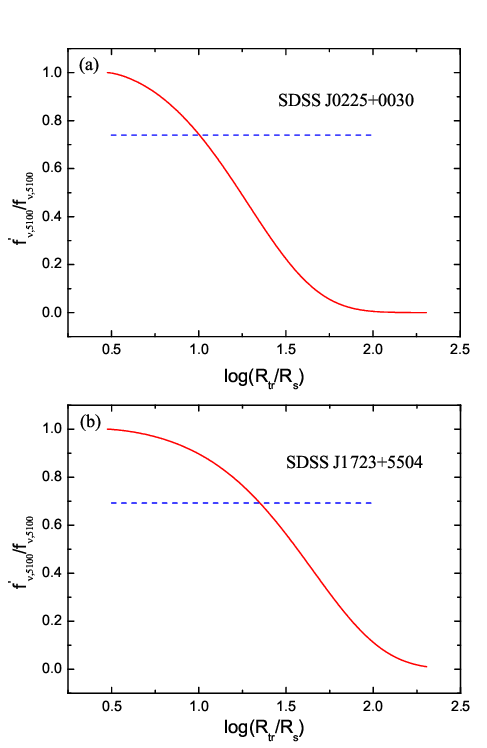}
\caption{Upper panel: variation of the ratio between the optical flux (5100 \AA) of the dim state and the bright state with the transition radius of a thin accretion disk for SDSS J0225+0030 (the red solid line), where the blue dashed line represents $f_{\rm {\nu,5100}}^{'}/f_{\rm {\nu,5100}}=74.13\%$. Lower panel: same as upper panel, but for the object SDSS J1723+5504, where the blue dashed line represents $f_{\rm {\nu,5100}}^{'}/f_{\rm {\nu,5100}}=69.2\%$.}\label{f2}
\end{figure}

\begin{figure}
\centering
\includegraphics[width=12cm]{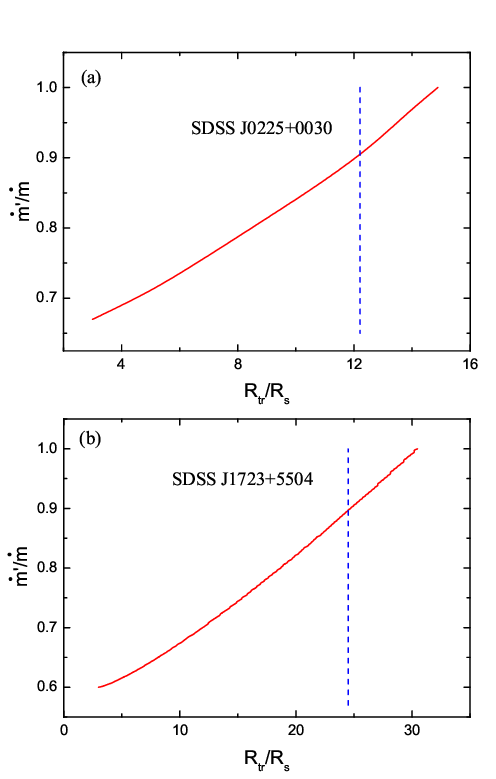}
\caption{Upper panel: variation of the mass accretion rate ratio between the dim state and bright state as a function of the transition radius $R_{\rm tr}$ for SDSS J0225+0030 (the red solid line), where the blue dashed line represents $R_{\rm tr}=12.21 R_{\rm s}$ ($t_{\rm cool}=t_{\rm tran}$). Lower panel: same as upper panel, but for the object SDSS J1723+5504, where the blue dashed line represents $R_{\rm tr}=24.5 R_{\rm s}$ ($t_{\rm cool}=t_{\rm tran}$).}\label{f3}
\end{figure}

\begin{figure}
\centering
\includegraphics[width=12cm]{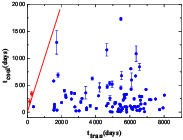}
\caption{Comparison of the observed transition timescales $t_{\rm tran}$ of CL AGNs with the cooling timescales $t_{\rm cool}$ given by our new model, where the solid red line represents
$t_{\rm tran}=t_{\rm cool}$. The two red dots with error bars represent SDSS J0225+0030 and SDSS J1723+5504, respectively.}\label{f4}
\end{figure}

The variation of optical flux at 5100 \AA{} with the inner radius of disk is investigated in Figure \ref{f2}. It is found that our new results can greatly shorten the cooling timescales. Assuming the flux variation from the dim state to the bright state is entirely caused by the transition from an ADAF to a thin disk in the inner disk region (with a constant mass accretion rate), $R_{\rm tr}$ will change from 78.1 $R_{\rm s}$ to 14.89 $R_{\rm s}$ for SDSS J0225+0030, reducing $t_{\rm cool}$ from 4112 days to 342 days (see Table 1 for details). Similarly, for SDSS J1723+5504, $R_{\rm tr}$ will change from 158.4 $R_{\rm s}$ to 30.48 $R_{\rm s}$, leading to the decrease of $t_{\rm cool}$ from 2299 days to 197 days. The calculated values of 342 and 197 days are still longer than the observational upper limits of 254 and 142 days in this case.


However, we assume that the mass accretion rate remains constant in Figure \ref{f2}. In fact, the mass accretion rate may increase during the state transition in physics \citep{2019ApJ...883...76R}. For a fixed observed ${L_{5100}^{'}}/L_{5100}$ ratio, if we assume that the mass accretion rate ($\dot{m}^{'}$) decreases significantly in the dim state, the reduction in optical flux can be largely attributed to the drop in $\dot{m}^{'}$, requiring a smaller change in $R_{\rm tr}$. In the limiting case where the flux reduction is entirely driven by a decrease in $\dot{m}^{'}$, the inferred $R_{\rm tr}$ will be equal to $3 R_{\rm s}$. In this case the thin disk extends to the Schwarzschild ISCO (innermost stable circular orbit), and no inner ADAF region is present.
Based on the observed ${L_{5100}^{'}}/L_{5100}$ ratio, we present the degeneracy between $\dot{m}'/\dot{m}$ and $R_{\rm tr}$ in Figure \ref{f3}. For the observed transition times of 254 (SDSS J0225+0030) and 142 (SDSS J1723+5504) days, the corresponding $R_{\rm tr}$ are 12.21 $R_{\rm s}$ and 24.5 $R_{\rm s}$, respectively.  Therefore, the parameter space defined by $3R_{\rm s}<R_{\rm tr} < 12.21 R_{\rm s}$ (implying $74.13\%<\dot{m}'/\dot{m} < 90.99\%$) can satisfy the condition $t_{\rm cool} < t_{\rm tran}$ for SDSS J0225+0030. Similarly, for SDSS J1723+5504, the corresponding parameter space is $3R_{\rm s}<R_{\rm tr} < 24.5 R_{\rm s}$ (with $69.2\%<\dot{m}'/\dot{m} < 89.74\%$).

The significant reduction of $R_{\rm tr}$ in our new model can be attributed to the 
large black hole mass ($\sim 10^9 M_{\odot}$) and the low Eddington-scaled mass accretion rate ($\sim 0.01$) in these two objects, resulting in a very low gas temperature in the inner disk region. The maximum temperatures in SDSS J0225+0030 and SDSS J1723+5504 are only $3.3\times10^4 {\rm K}$ and $5.5\times10^4 {\rm K}$, respectively. Therefore, the optical flux at $5100$ \AA{} (corresponding to a temperature $5700$ $ K$) can be greatly decreased even if the transition radius $R_{\rm tr}$ is quite small. In addition, a Schwarzschild black hole is adopted in our new model. We suggest that the results will be qualitatively the same for a Kerr black hole because the effects of a Kerr black hole mainly manifest in the inner disk region, which has a higher temperature and contributes little to the optical flux around 5100 \AA. To further test our new model, we apply it to the full sample of LC2025, whose physical properties are summarized in Table 2. We exclude two objects from \citet{2025ApJ...980...91Y}, SDSSJ1327+4025 and SDSSJ1341-0049, as their continuum luminosities decline with the emergence of broad emission lines, because our model can only be applied to the objects with increasing luminosities as they brighten. These transitions are likely triggered by the dust obscuration instead of an accretion disk state transition.
As shown in Figure \ref{f4}, the cooling timescales predicted by our model do not exceed the observational transition-timescale constraints.

\begin{table*}[!htbp]
\caption{The physical properties of SDSSJ0225+0030 and SDSSJ1723+5504.}
\setlength{\tabcolsep}{7pt} 
\begin{tabular*}{\linewidth}{@{\extracolsep{\fill}} ccccccccccc @{\extracolsep{\fill}}}
\hline

{Name} & {z} & {log$m_{\rm bh}$} & {log$L_{5100}^{'}$} & {log$L_{5100}$} &  {$\frac{L_{5100}^{'}}{L_{5100}}$} & {$t_{\rm tran}$} & {$R_{\rm {tr,0}}$} & {$t_{\rm cool,0}$} & {$R_{\rm {tr}}$} & {$t_{\rm cool}$} \\

{(1)} &  {(2)} &  {(3)} &  {(4)} & 
 {(5)} &  {(6)} &  {(7)} &
 {(8)} &  {(9)} &  {(10)} & {(11)} \\
\hline
SDSSJ0225+0030   &  0.504  &	9.27${\pm0.09}$   &  44.16  & 44.29 & 	 	74.1\%    &  254  &  	78.1  &	 4112   &  	 	14.89${\pm1.20}^*$    &  342${\pm41}$     \\
 	
SDSSJ1723+5504   &  0.295 &		8.80${\pm0.15}$    &  44.09 & 44.25  &		69.2\%	   &  142 & 158.4 &	2299   &	30.48${\pm4.91}$  &	197${\pm47}$    \\

\hline
\multicolumn{11}{p{\textwidth}} {Notes: Col. (1): Source name. Col. (2): Redshift. Col. (3): Black hole mass $m_{\rm bh}=M_{\rm BH}/M_{\odot}$. Col. (4): Optical continuum luminosity at 5100 \AA{} in the dim state ($L_{5100}^{'}=4\pi D^2 f_{\nu,5100}^{'} $). Col. (5): Optical continuum luminosity at 5100 \AA{} in the bright state. Col. (6): Ratio of $L_{5100}^{'}/L_{5100}$  Col. (7): Observed timescale of the object. Col.(8): Transition radius given by LC2025. Col. (9): Cooling timescale given by LC2025. Col.(10): Transition radius given by our new model assuming that the optical flux variation originates from the decrease of transition radius only. Col.(11): Cooling timescale corresponding to the transition radius in Col. (10). }\\
\multicolumn{11}{p{\textwidth}} {$^*$: Here, the uncertainties for $R_{\rm tr}$ and $t_{\rm cool}$ represent solely the formal propagated errors from the spectral fitting; the uncertainties related to the model parameters (e.g., the viscosity parameter $\alpha$, black hole spin $a_{*}$, and $f_{\rm adv}$) are not included (see LC2025 for details).  
}  \\
\end{tabular*}

\end{table*}

\begin{table*}[!htbp]
\caption{The physical properties of the LC2025 sample.}
\setlength{\tabcolsep}{3pt} 
\begin{tabular*}{\linewidth}{@{\extracolsep{\fill}} lllllllcllll @{\extracolsep{\fill}}}
\hline

\multicolumn{1}{c}{Name} & \multicolumn{1}{c}{z} & \multicolumn{1}{c}{log$m_{\rm bh}$} & \multicolumn{1}{c}{log$L_{\lambda}^{'}$$^*$} & \multicolumn{1}{c}{log$L_{\lambda}$} &  \multicolumn{1}{c}{$\frac{L_{\lambda}^{'}}{L_{\lambda}}$} & \multicolumn{1}{c}{$t_{\rm tran}$} & \multicolumn{1}{c}{Ref.}  & \multicolumn{1}{c}{$R_{\rm {tr,0}}$} & \multicolumn{1}{c}{$t_{\rm cool,0}$}  & \multicolumn{1}{c}{$R_{\rm tr}$} & \multicolumn{1}{c}{$t_{\rm cool}$} \\

\multicolumn{1}{c}{(1)} &  \multicolumn{1}{c}{(2)} &  \multicolumn{1}{c}{(3)} &  \multicolumn{1}{c}{(4)} & 
 \multicolumn{1}{c}{(5)} &  \multicolumn{1}{c}{(6)} &  \multicolumn{1}{c}{(7)} &
 \multicolumn{1}{c}{(8)} &  \multicolumn{1}{c}{(9)} &  \multicolumn{1}{c}{(10)} & \multicolumn{1}{c}{(11)} & \multicolumn{1}{c}{(12)} \\
\hline

SDSSJ0002-0027 & 0.291 & $8.94\pm0.02$ & $43.95$ & $44.12$ & $0.68$ & 2260 & 1 & 115.40 & 2764 & $24.73\pm0.56$ & $274\pm9$ \\
SDSSJ0023+0035 & 0.422 & $9.17\pm0.04$ & $44.55\pm0.01$ & $44.73$ & $0.66\pm0.02$ & 2577 & 1 & 130.00 & 3507 & $28.43\pm1.56$ & $359\pm30$ \\
SDSSJ0040+1609 & 0.294 & $7.85\pm0.04$ & $43.48$ & $43.71$ & $0.6$ & 6412 & 3 & 402.60 & 813 & $48.65\pm1.78$ & $34\pm2$ \\
SDSSJ0144+3140 & 0.124 & $7.66\pm0.03$ & $44.0\pm0.01$ & $44.28\pm0.01$ & $0.52\pm0.02$ & 363 & 2 & 470.50 & 663 & $134.24\pm10.94$ & $101\pm12$ \\

\hline

\multicolumn{12}{p{\textwidth}} {Notes: Same as Table 1 except for Col. (8), where (1), (2), and (3) represent \citet{2024ApJS..272...13P}, \citet{2025ApJ...986..160D}, and \citet{2025ApJ...980...91Y}, respectively.}\\

\multicolumn{12}{p{\textwidth}} {$^*$: The optical continuum luminosities are provided at 5100 \AA{} for the objects from \citet{2024ApJS..272...13P} and \citet{2025ApJ...986..160D}, while \citet{2025ApJ...980...91Y} reported the continuum luminosity at 3000 \AA{}. }  \\

\multicolumn{12}{p{\textwidth}} {$^{**}$: This table is available in its entirety in the online article.  }  \\

\end{tabular*}

\end{table*}

\section{Summary and Discussion} \label{discussion}

The LC2025 model can qualitatively explain the observational timescales of turn-on CL AGNs, except for SDSS J0225+0030 and SDSS J1723+5504. We investigate this problem by considering two other theoretical models. Our new results mainly include:

1), the effect of a large-scale poloidal magnetic field is included to reduce the viscous timescales by improving the gas radial velocity. However, the viscous timescales in these two objects, though decreasing about three orders of magnitude, are still too long compared to the observational transition timescales of these two objects due to their large black hole mass and small mass accretion rate. 

2), we revise the model of LC2025 by improving the method to recalculate the transition radius $R_{\rm tr}$ based on the variation of optical flux, which can significantly enhance the accuracy of our results. It is found that the cooling timescales provided by our new model are comparable to the observed transition timescales of the SDSS J0225+0030 and SDSS J1723+5504.

3), we further apply our new model to the full LC2025 sample (except for SDSSJ1327+4025 and SDSSJ1341-0049). It is found that the cooling timescales provided by our new model are comparable to or shorter than the observed transition timescales in all objects.

Both the large-scale poloidal magnetic field and the large-scale toroidal magnetic field can improve the radial velocity of accretion disk. We investigate the effect of a large-scale poloidal magnetic field on the viscous timescales of CL AGNs in this work only. However, as shown in previous works \citep{2019MNRAS.483L..17Dexter,2023ApJ...958..146W}, though the large-scale toroidal magnetic field can improve the radial velocity of disk by increasing the magnetic pressure, the viscous timescales can only be shortened by about 3-4 orders of magnitude. This is insufficient to account for the observed timescale either. Furthermore, the observational radio emission in SDSS J1723+5504 is below the detection limit of FIRST (no data for SDSS J0225+0030, see
https://vizier.cds.unistra.fr/viz-bin/VizieR), contradicting the prediction when the large-scale poloidal/toroidal magnetic field is strong.

\section* {Acknowledgements}
We appreciate the referee's very valuable and helpful comments, which helped to improve and clarify this manuscript. This work is supported by the National Key R\&D Program of China
(Grant No. 2023YFA1607903, 2025YFA1614102), the NSFC (grants 12273089, 12533005, 12233007, 12347103, and 12361131579), and the science research grants from the China Manned Space Project with CMS-CSST-2025-A07.

\bibliography{CLAGN}
\bibliographystyle{aasjournal}

\appendix
\section{Effects of a large-scale poloidal magnetic field}\label{model I}

\begin{figure}
\centering
\includegraphics[width=12cm]{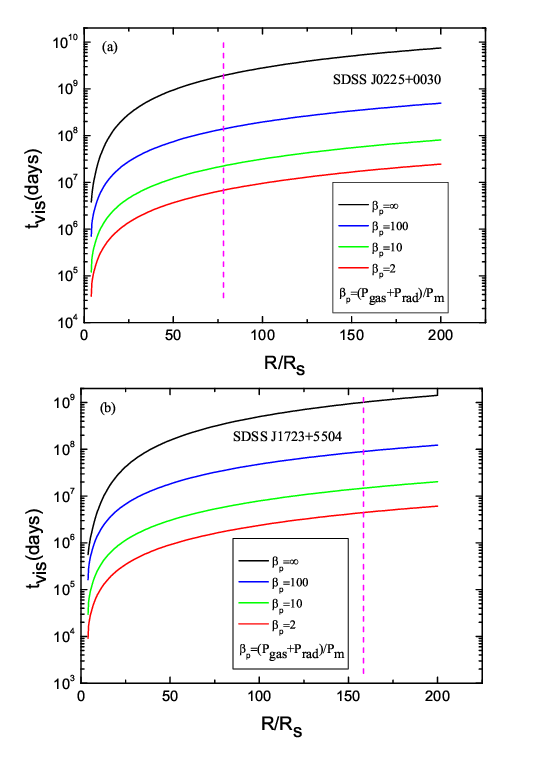}
\caption{Upper panel: viscous timescale as functions of radius for SDSS J0225+0030, where the black hole mass $M_{\rm bh}=1.86\times 10^9 M_{\odot}$, viscosity parameter $\alpha=0.1$ and mass accretion rate $\log\dot{m}=-2.15$ ($\dot{m}=\dot{M}/\dot{M}_{\rm Edd}$, where $\dot{M}$ and $\dot{M}_{\rm Edd}$ are the Eddington scaled mass accretion rate and Eddington mass accretion rate, respectively) are adopted through observations \citep{2016MNRAS.457..389M}. The black, blue, green and red lines correspond with $\beta_{\rm p}=\infty$, $\beta_{\rm p}=100$,$\beta_{\rm p}=10$ and $\beta_{\rm p}=2$, respectively. The magenta dashed line represents the transition radius $R_{\rm tr}=78.1 R_{\rm s}$ (see \citealt{2025ApJ...988..207L} for details). Lower panel: same as upper panel, but for the object SDSS J1723+5504, where the black hole mass $M_{\rm bh}=6.31\times 10^8 M_{\odot}$, viscosity parameter $\alpha=0.1$ and mass accretion rate $\log\dot{m}=-1.73$ are adopted \citep{2021A&A...650A..33P}. The magenta dashed line represents the transition radius $R_{\rm tr}=158.4 R_{\rm s}$. Note: even though $R_{\rm tr}$ given by the revised model can decrease to 10.83 $R_{\rm s}$ and 21.52 $R_{\rm s}$ for SDSS J0225+0030 and SDSS J1723+5504, respectively, the corresponding viscous timescales are still far longer than one year.}\label{f1}
\end{figure}

The cooling timescales of an ADAF in these two objects are about one order of magnitude longer than the observed transition timescales (LC2025). As mentioned above, the presence of a large-scale poloidal magnetic field can significantly improve the radial velocity of accretion disk by transferring the angular momentum through outflows. Therefore, we will take the effect of a large-scale magnetic field into account to explore this issue first. Following the formalisms of \citet{2014ApJ...786....6L}, the basic equations of a thin disk with a large-scale magnetic field around a Kerr black hole can be written as:
\begin{equation}
\frac{d}{dR}(2\pi \Delta^{1/2} \Sigma v_{\rm R}/\gamma_\phi)+4\pi R \dot{m}_{\rm w}=0, \label{continuity}
\end{equation}
\begin{equation}
\frac{\gamma_\phi A M}{R^4 \Delta}\frac{(\Omega-\Omega_{\rm K}
^+)(\Omega-\Omega_{\rm K} ^-)}{\Omega_{\rm K} ^+ \Omega_{\rm K}
^-}+g_{\rm m}=0, \label{momentum}
\end{equation}
\begin{equation}
-\frac{\dot{M}}{2\pi} \frac{dL}{dR} + \frac{d}{dR}(R
W^R_\phi)+T_{\rm m}R=0, \label{angular}
\end{equation}
\begin{equation}
\nu \Sigma \frac{\gamma_{\phi}^4 A^2}{R^6}\left(
\frac{d\Omega}{dR}\right)^2= \frac{16acT^4}{3\bar{\kappa}\Sigma}
\label{energy},
\end{equation}
where $g_{\rm m}=B_{\rm R}B_{\rm Z}/2\pi\Sigma$ and $T_{\rm m}=B_{\rm P}B_{\rm \phi}R/2\pi$ are the radial magnetic force and magnetic torque, respectively, and the other physical parameters have common meanings. The black hole spin $a_*=0$ is adopted for simplicity in this work. The four variables $\rho$, $V_{\rm R}$, $\Omega$, and $T$ in equations above can be numerically solved when the black hole mass ($M_{\rm bh}$), mass accretion rate ($\dot{M}$), viscosity parameter ($\alpha$), and $\beta_{\rm p}=(P_{\rm gas}+P_{\rm rad})/P_{\rm m}$ ($P_{\rm gas}$, $P_{\rm rad}$, and $P_{\rm m}$ are the gas pressure, radiative pressure and the magnetic pressure, respectively) are adopted. As \citet{2014ApJ...786....6L}, we employ the Newton-Raphson method to solve the equations from $R_{\rm ISCO}(=GM/c^2)$ to the outer radius of the accretion disk (see section 3.2 of \citealt{2014ApJ...786....6L} for details).

The effects of a large-scale poloidal magnetic field on the viscous timescales of a thin accretion disk are investigated in Figure \ref{f1}, where Figure \ref{f1}(a) and Figure \ref{f1}(b) correspond to SDSS J0225+0030 and SDSS J1723+5504, respectively, and the magenta lines represent the transition radius $R_{\rm tr}$ between the inner ADAF and the outer thin disk given by LC2025. It is found that viscous timescales of a thin disk with strong magnetic field ($\beta_{\rm p}=2$) can be reduced about three orders of magnitude compared to those of a standard thin disk. However, even the shortest viscous timescales in both objects ($\sim 10^6$ days) are still several orders of magnitude longer than the observed transition timescales, due to their large black hole masses ($\sim 10^9 M_{\odot}$) and small mass accretion rates ($\sim 0.01$). Therefore, the effects of a large-scale poloidal magnetic field are inadequate to explain the observed transition timescales in SDSS J0225+0030 and SDSS J1723+5504.

\end{document}